# STABILITY OF BINARY ON HIGHLY ECCENTRIC ORBIT

Rosaev A.E.
OAO NPC "Nedra" Jaroslavl, Russia, *hegem@mail.ru*

It is known, that possibility to capture on a satellite orbit is decreased with primary orbit's eccentricity increasing. It means, that satellites of planet on eccentric orbit are less stable. We study problem by analytical and numeric methods. A difference between case of infinitesimal mass satellite and true binary system is shown. The dependence of orbital energy on external eccentricity is derived for binary system.

In result of our studying, we can conclude, that extrasolar planets may have significantly smaller satellites, than planets in our solar system. Moreover, dependence on eccentricity makes restriction on giant planets migration in our Solar System. The application to a flight dynamics and mission design is discussed.

## 1. Introduction

Problem of stability satellite of planet at highly eccentric heliocentric orbit have a number of various applications. First, considered task is related with NEO origin problem. In all cases, transfer from main belt to NEO region followed eccentricity growth. On the other hand, many asteroids that make close encounters with terrestrial planets are in a binary configuration. Secondly: extrasolar planets have a large eccentricity. Can they have satellites on stable orbits? Third application: Large planets migration in a solar system. In particular, migration process followed by significant eccentricity increasing. Fourth application is to new space mission preparation.

In according with result [1], possibility to capture on a satellite orbit is decreased with primary orbit's eccentricity increasing. But the planetary migration process is followed by significant eccentricity increasing. In this paper we show, that planet on eccentric orbit must have smaller number than on circular ones. This process decreases the stability of satellites. It means, that primordial satellite system may be destroyed after the origin. Here we use direct mutual energy estimations. The description of our numeric method and very preliminary results are given in [2,3].

## 2. Model description

Let to consider planar three body problem (RTBP). Geometry of problem is given in fig.1.

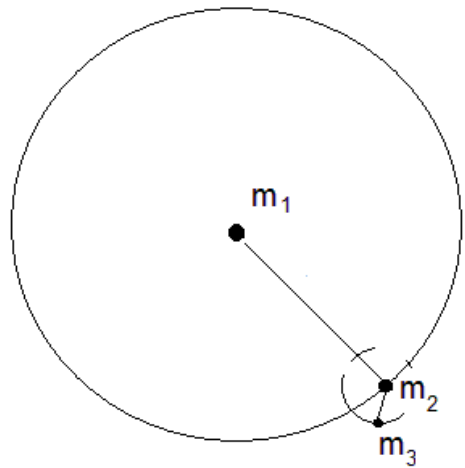

Fig.1. Problem geometry.

Main notations are: $m_1$ – main body or reduced mass in barycentre, $m_2$ – secondary mass $m_3$ – infinitesimal test particle, $f$ – is a gravity constant, $V_i$ – respective velocities, $\Delta_{ij}$ - distances between points. Hamiltonian for this problem is:

$$H = 1/2 \sum m_i V_i^2 - 1/2 f \sum_{j=0}^{n-1} \sum_{i=0}^{n-1} \frac{m_j m_i}{\Delta_{ij}} \tag{1}$$

Let $m_1=1, V_1=0$.

$$H = 1/2 m_2 V_2^2 + 1/2 m_3 V_3^2 - f \frac{m_2}{\Delta_{12}} - f \frac{m_3}{\Delta_{13}} - f \frac{m_2 m_3}{\Delta_{23}} \tag{2}$$

Denote $V_{23}$ - velocity of $m_3$ relative the second primary:

$$V_3^2 = V_2^2 + V_{23}^2 - 2 V_2 V_{23} \cos(V_2, V_{23}) \tag{3}$$

Energy integral becomes:

$$H = 1/2(m_2 + m_3) V_2^2 + 1/2 m_3 V_{23}^2 - m_3 V_2 V_{23} \cos(V_2, V_{23}) - f \frac{m_2}{\Delta_{12}} - f \frac{m_3}{\Delta_{13}} - f \frac{m_2 m_3}{\Delta_{23}} \tag{4}$$

Denote $h_{23}$ - energy of $m_3$ relative the second primary.

$$h_{23} = 1/2 m_3 V_{23}^2 - f \frac{m_2 m_3}{\Delta_{23}} = H - 1/2(m_2 + m_3) V_2^2 + m_3 V_2 V_{23} \cos(V_2, V_{23}) + f \frac{m_2}{\Delta_{12}} + f \frac{m_3}{\Delta_{13}} \tag{5}$$

Here:

$$\Delta_{13} = \sqrt{r_{12}^2 + r_{23}^2 - 2 r_{12} r_{23} \cos(\alpha)}, \quad \Delta_{12} = r_{12} = \frac{1 - e_{12}^2}{1 + e_{12} \cos \vartheta_{12}} \tag{6}$$

and $\cos \alpha \approx \cos(V_2, V_{23})$.

Ad definition. (Belbruno [4]) $P_3$ is ballistically captured at $P_2$ at time $t$ if the two body Kepler energy of $P_3$ with respect in $P_2$-centred inertial coordinates:

$$h_{23}(\mathbf{X}, \dot{\mathbf{X}}) = \frac{1}{2} \dot{\mathbf{X}} - \frac{\mu}{|\mathbf{X}|} \leq 0, \quad 0 \leq \mu < 1/2$$

for a solution $\phi(t) = (\mathbf{X}, \dot{\mathbf{X}})$ of the elliptic restricted problem relative to $P_2$, $|X|>0$

### 3. Small satellite around NEA

Let to consider a limit case $m_3=0$. Hill sphere method is well known for studying stability of asteroid small satellite. If the mass of the smaller body (e.g. asteroid primary) is $m$, and it orbits a heavier body (e.g. Sun) of mass $M$ with a semi-major axis $a$ and an eccentricity of $e$, then the radius $r$ of the Hill sphere for the smaller body (e.g. asteroid primary) is, approximately[5]:

$$r \approx a(1-e) \sqrt[3]{\frac{m}{3M}} \tag{7}$$

It means, that the stability zone for asteroid satellite is decrease with eccentricity increasing. Moreover, perturbations of satellite orbit of binary eccentric are strong

If eccentricity decrease or constant, $\Delta e(t) \leq 0$, then energy negative $m_3$ forever on satellite orbit around $m_2$. However, if eccentricity increase with time, $\Delta e(t) > 0$, then energy can become positive, and binary is destroyed. During migration from main belt to NEA region, eccentricity increase about 0.3.

Actually, gravitational spheres and related zero-velocity (Hill sphere) have non-spherical form. They have complex topology, depends from masses and eccentricities of primary bodies. Asymmetry always take place in direction toward to primary, action sphere has minimal, in opposite direction – maximal radius.

$$\rho = \frac{R_{p,s}\sqrt{\frac{m}{M}}}{\left(1 - \cos\varphi\sqrt{\frac{m}{M}}\right)} \qquad (8)$$

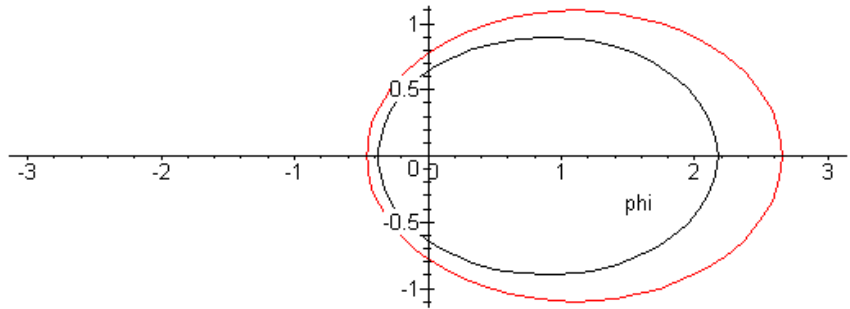

Fig.2. The example of the action sphere projection at perihelia and aphelia

**4. Binary case**

Lets $m_3 \neq 0$. Return to equation (5):

$$h_{23} = H - 1/2(m_2 + m_3)V_2^2 + m_3 V_2 V_{23} \cos(V_2, V_{23}) + f\frac{m_2}{r_{12}} + f\frac{m_3}{\Delta_{13}} \qquad (9)$$

Let $\cos\alpha \approx \pm\cos(V_2, V_{23})$. Sign + is valid for retrograde, sign – for prograde orbits (Fig. 3).

$$\Delta_{13} = \sqrt{r_{12}^2 + r_{23}^2 - 2r_{12}r_{23}\cos(\alpha)} \qquad (10)$$

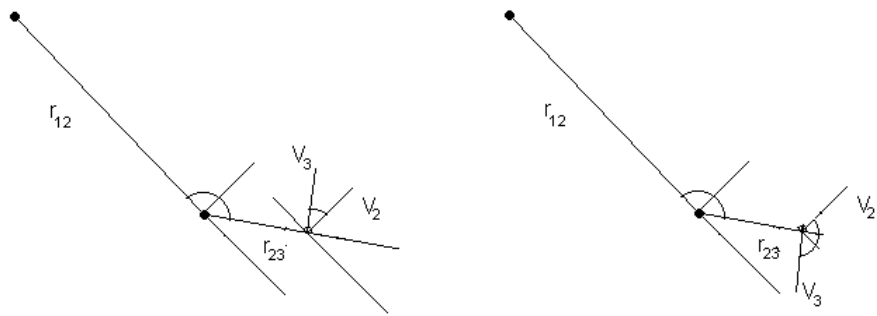

Fig.3. On definition of angle α for prograde (left) and retrograde (right) orbits.

In our notations (M=1):

$$V_2 = \sqrt{fM \frac{2(1+e_{12}\cos\vartheta)-(1-e_{12}^2)}{a_{12}(1-e_{12}^2)}} = \sqrt{fM \frac{1+2e_{12}\cos\vartheta+e_{12}^2}{a_{12}(1-e_{12}^2)}} \qquad (11)$$

At pericenter orbit of $m_3$ around $m_2$:

$$V_{23} = \sqrt{fm_2 \frac{2(1+e_{23}\cos\theta)-(1-e_{23}^2)}{a_{23}(1-e_{23}^2)}} = \sqrt{fm_2 \frac{1+2e_{23}+e_{23}^2}{a_{23}(1-e_{23}^2)}} = \sqrt{fm_2 \frac{1+e_{23}}{a_{23}(1-e_{23})}} \qquad (12)$$

After substitution $V_{23}$:

$$h_{23} = H - 1/2(m_2+m_3)V_2^2 + \frac{fm_2}{r_{12}} \pm m_3 V_2 \sqrt{\frac{fm_2(1+e_{23})}{r_{23}(1-e_{23})}} \cos\alpha + \frac{fm_3}{\sqrt{r_{12}^2 + r_{23}^2 - 2r_{23}r_{12}\cos\alpha}} \qquad (13)$$

After using expression (8):

$$h_{23} = H + 1/2 fm_2 \frac{1}{a_{12}} + m_3 V_2 \left[ -V_2/2 \pm \sqrt{\frac{fm_2(1+e_{23})}{r_{23}(1-e_{23})}} \cos\alpha \right] + \frac{fm_3}{\sqrt{r_{12}^2 + r_{23}^2 - 2r_{23}r_{12}\cos\alpha}} \qquad (14)$$

Here and below sign "+" for prograde and sign "−" for retrograde orbit of satellite.
After substitution $V_2$ and simplification:

$$h_{23} = H + 1/2 m_2 \frac{1}{a_{12}} - m_3 \left[ \frac{(1+2e_{12}\cos\vartheta_{12}+e_{12}^2)}{2a_{12}(1-e_{12}^2)} \pm \sqrt{\frac{m_2(1+2e_{12}\cos\vartheta_{12}+e_{12}^2)}{a_{12}(1-e_{12}^2)r_{23}}} (1+e_{23})\cos\alpha \right] +$$

$$+ \frac{m_3}{\sqrt{r_{12}^2 + r_{23}^2 - 2r_{23}r_{12}\cos\alpha}}$$

$$(15)$$

We search for dependence $h_{23}$ on $e_{12}$. Apply Legendre expansion to last term:

$$\frac{m_3}{\sqrt{r_{12}^2 + r_{23}^2 - 2r_{23}r_{12}\cos\alpha}} = \frac{m_3}{r_{12}} \left[ 1 + \frac{r_{23}}{r_{12}}\cos\alpha + 1/2 \left[\frac{r_{23}}{r_{12}}\right]^2 (3\cos^2\alpha - 1) + ... \right] \qquad (16)$$

At pericenter orbit of $m_2$:

$$h_{23} = H + (m_2+2m_3)\frac{1}{2a_{12}} - \frac{m_3}{a_{12}} \left[ \frac{(1+e_{12})}{2(1-e_{12})} + \left( \frac{r_{23}}{r_{12}} \pm \sqrt{\frac{m_2 a_{12}(1+e_{12})(1+e_{23})}{r_{23}(1-e_{12})(1-e_{23})}} \right) \cos\alpha \right] +$$

$$+ 1/2 \frac{m_3}{a_{12}} \left[\frac{r_{23}}{r_{12}}\right]^2 (3\cos^2\alpha - 1) + 1/2 \frac{m_3}{a_{12}} \left[\frac{r_{23}}{r_{12}}\right]^3 (5\cos^3\alpha - 3\cos\alpha) + ... \qquad (17)$$

Restrict by main terms and use $\dfrac{f}{a_{12}} = \dfrac{2f}{r_{12}} - V_2^2$ :

$$h_{23} = H + 1/2(m_2 + m_3)\frac{1}{a_{12}} - \frac{m_3}{a_{12}}\left[\pm\sqrt{\frac{m_2 a_{12}(1+e_{12})(1+e_{23})}{r_{23}(1-e_{12})(1-e_{23})}} + \frac{r_{23}}{r_{12}}\right]\cos\alpha \tag{18}$$

Total energy $H$ consists of energy $m_3$ and $m_2$ on orbital around $M$ and energy $m_3$ on orbit around $m_2$. As for case infinitesimal mass of satellite, energy is bounded, so we need to add $E_b$ related with Hill sphere radius:

$$H = -(m_2 + m_3)\frac{1}{2a_{12}} - \frac{m_2 m_3}{2r_{23}} + E_b \tag{19}$$

It is naturally to accept:

$$E_b = \pm\frac{m_3}{a_{12}}\sqrt{\frac{m_2 a_{12}(1+e_{12})(1+e_{23})}{r_h(1-e_{12})(1-e_{23})}} + \frac{m_3}{a_{12}}\frac{r_h}{r_{12}} \tag{20}$$

Finally we can rewrite (18):

$$h_{23} = -\frac{m_2 m_3}{2r_{23}} + E_b - \frac{m_3}{a_{12}}\left[\pm\sqrt{\frac{m_2 a_{12}(1+e_{12})(1+e_{23})}{r_{23}(1-e_{12})(1-e_{23})}} + \frac{r_{23}}{a_{12}}\right]\cos\alpha \tag{21}$$

The conditions when $m_3$ neglected is followed from equation (20):

$$\sqrt{\frac{m_2 a_{12}}{m_3 r_{23}}} \ll 1 \tag{22}$$

It means, that for most real asteroid binaries mass of satellite can be neglected. In contrary, mass of planet satellites need to take into account.

Accept $a_{12} = 1$, $m_2 = 1$. The value of $\cos\alpha_{escape} \approx 0.1$ is estimated empirically from stability of outer Jovian satellites. Condition of escape:

$$h_{23} = -\frac{1}{2r_{23}} - \frac{1}{r_b} + 0.1\left[\sqrt{\frac{m_2(1+e_{12})}{m_1 r_{23}(1-e_{12})}} + \frac{r_{23}}{1}\right] \tag{23}$$

Actually, capture/escape possibility is proportional to ratio

$$\frac{1}{r_b} \Big/ \sqrt{\frac{m_2(1+e_{12})}{m_1 r_{23}(1-e_{12})}} \tag{24}$$

On a first look, the escape is more easy at large $m_2/m_1$ mass ratio. But after take into account:

$$r_b \approx a(1-e)\sqrt[3]{\frac{m_2}{3m_1}} \tag{25}$$

It is evident, that asteroidal binary is not more stable, than large planet satellite. Graph for Jovian system mass ratio $m_2/m_1 = 0.001$ and different $m_3$ is given at fig.4-5.

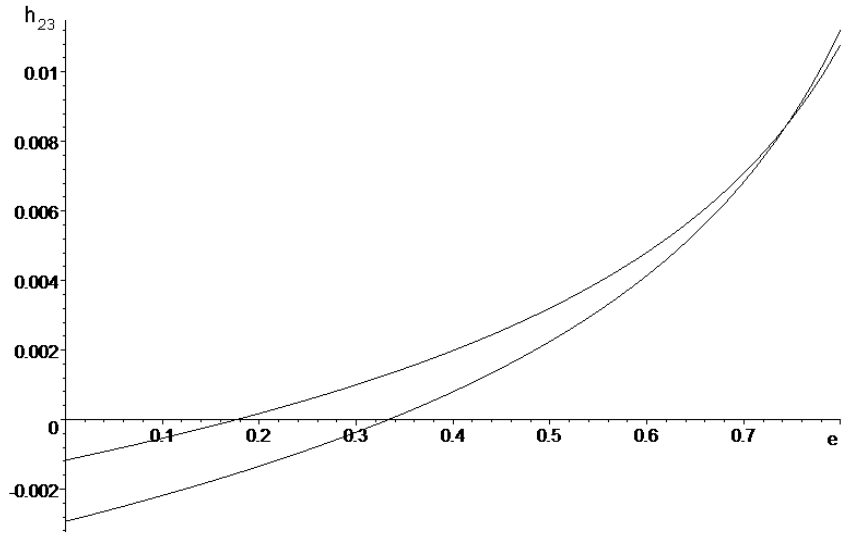

Fig.4. Relative energy as a function of the external eccentricity $e_{12}$. The bottom curve for case $m_3 = 0$

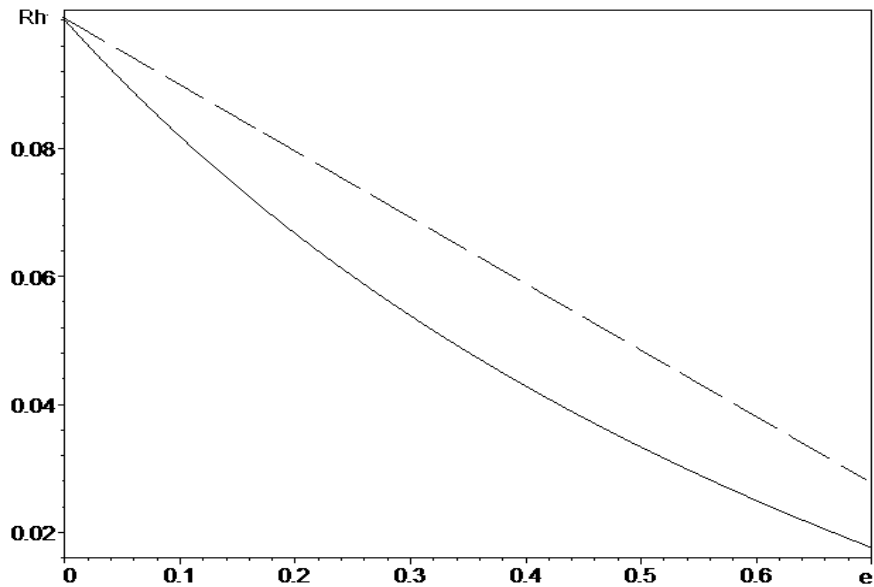

Fig.5. Effective Hill radius as a function of external eccentricity $e_{12}$.
Dashed line – for $m_3 = 0$.

## 5. Applications

**Near Earth Objects.** All of NEA binaries with well known orbits are really a system with large mass and small satellite, not a binary with compatible mass. Most of them have quasi-circular regular orbits deeply inside Hill sphere, some of them are on synchronous orbits [6]. In means, that lifetime of considered NEO satellites is sufficient to regularize their orbit. Moreover, time since last close encounter with Earth is not small.

Table 1

NEA binaries with most well established orbits[6]

| | $a$ | $e$ | $m_3/m_2$ | $a_{bin}$ | $R_h$ |
|---|---|---|---|---|---|
| (65803) Didymos | 1.64434 | 0.38401 | 0.0066 | 1.18 | 209.8 |
| (185851)2000DP$_{107}$ | 1.36612 | 0.37675 | 0.05 | 2.62 | 61.2 |
| (66391) 1999 KW$_4$ | 0.64229 | 0.68839 | 0.057 | 2.55 | 68.4 |
| (276049)2002 CE$_{26}$ | 2.23317 | 0.55963 | 0.0006 | 4.7 | 705.8 |
| (164121) 2003 YT$_1$ | 1.10997 | 0.29187 | 0.007 | 3.4 | 249.7 |
| (35107) 1991 VH | 1.13635 | 0.14374 | 0.064 | 3.2 | 279.5 |
| (153591)2001SN$_{263}$ | 1.98531 | 0.47767 | 0.010 | 16.633 | 567.8 |
| | | | 0.026 | 3.804 | |
| (136617) 1994 CC | 1.63899 | 0.41755 | 0.022 | 1.729 | 156.2 |
| | | | 0.0035 | 6.130 | |

It means, that problem of NEA binaries origin remain unresolved. They can origin as binary in main belt, as well as to be a product of catastrophic breakups in situ.

**Flight dynamics.** The possible use of results in flight dynamic include: 1) launch Cosmic Mission to elliptic resonance heliocentric orbit; 2) separate spacecraft into two apparatus on orbits, close to Hill radius; 3) use described effect of resonance heliocentric eccentricity growth to final separation of apparatus trajectories. In results we have two spacecrafts with close trajectories with different scientific programs (orbital modulus and landing modulus for example). Variation of eccentricity on resonance heliocentric orbit can reach value about 0.1. It reduces effective Hill radius from 0.08 to 0.07 (i.e. about 12%). It is possible fuel economy at spacecrafts trajectory correction impulse. Maximal economy obtained in case comparable masses of spacecrafts and low eccentricity of initial heliocentric orbit.

**Planetary migration.** Each planet, which orbital eccentricity is increased during migration, loss their satellite system. It may be actually true for Mercury and Venus case both for case hypothetic solar system fifth planet **[7]**.

**Results and discussion**

Now we discuss equation (21) and give some possible applications. Immediately, we can conclude, that retrograde orbit is more stable, then prograde. Suppose **m$_3$** on circular orbit around *m$_2$* and initial orbit of *m$_2$* circular.

It is evident, that relative energy decrease, when eccentricity of binary increase. It means, that binary system become unstable and destroyed. It will be take place at any scenario – at fixed q or at fixed a. In last case destruction of binary is more rapidly. Retrograde orbits in binary are more stable and in most cases can hold during migration process, followed eccentricity growth. Infinitesimal mass satellite is rather more stable than binary system with comparable masses**.**

The possible use of results in flight dynamic can provide fuel economy for some new missions.

Another interesting application of our result about instability satellite system at large eccentricity is a planetary migration. Each planet, which orbital eccentricity is increased during migration, loss their satellite system. It may be actually true for Mercury and Venus case both for case hypothetic solar system fifth planet **[7]**.

Here we not consider close encounters binary with planets. However, in case relatively slow eccentricity growth due to resonant perturbation, many of binary, which born in main belt destroyed before become NEA.